# A Combined Model for Noise Reduction of Lung Sound Signals Based on Empirical Mode Decomposition and Artificial Neural Network

Mozhde Firoozi Pouyani[1], Mansour Vali[2]*, Mohammad Amin Ghasemi [3]

[1,2] Department of Electrical Engineering, K. N. Toosi University of Technology, Tehran, Iran

[3] Department of Electrical and Computer Engineering, Tarbiat Modares University, Tehran, Iran

**Abstract**

Computer analysis of Lung Sound (LS) signals has been proposed in recent years as a tool to analyze the lungs' status but there have always been main challenges, including the contamination of LS with environmental noises, which come from different sources of unlike intensities. One of the common methods in noise reduction of LS signals is based on thresholding on Discrete Wavelet Transform (DWT) coefficients or Empirical Mode Decomposition (EMD) of the signal, however in these methods, it is necessary to calculate the SNR value to determine the appropriate threshold for noise removal. To solve this problem, a combined model based on EMD and Artificial Neural Network (ANN) trained with different SNRs (0, 5, 10, 15, and 20dB) is proposed in this research. The model can denoise white and pink noises in the range of -2 to 20dB without thresholding or even estimating SNR, and at the same time, keep the main content of the LS signal well. The proposed method is also compared with the EMD-custom method, and the results obtained from the SNR, and fit criteria indicate the absolute superiority of the proposed method. For example, at SNR = 0dB, the combined method can improve the SNR by 9.41 and 8.23dB for white and pink noises, respectively, while the corresponding values are respectively 5.89 and 4.31dB for the EMD-Custom method.

## 1. Introduction

Respiratory diseases are one of the leading causes of early death. According to the World Health Organization (WHO), lung diseases cause more than one million deaths worldwide, mainly due to the lack of timely diagnosis and progression of the disease [1]. The analysis of respiratory sounds by a specialist can lead to the diagnosis of symptoms or the disease itself. Stethoscope-based auscultation techniques make it possible to examine the respiratory system quickly and without side effects by listening to the sound of breathing. The mortality rate due to pulmonary disorders can be reduced by timely detection of pathological changes. Most doctors now have complete confidence in this method. However, the fact is that the lung sound with the help of a mechanical and even a digital stethoscope is not very reliable because the conclusion about the existence of disorders in the patient's lungs depends entirely on the ability and experience of the doctor [2]. Therefore, many researchers have considered the computer analysis of lung sounds in recent years [3].

---

[1] M.pouyani@ieee.org
*Corresponding Author:[2] Mansour.vali@eetd.kntu.ac.ir
[3] Aminghasemi@live.com

The computerized lung sound analysis involves recording LS signals by electrical devices, allowing researchers and physicians to store these signals and use machine learning and pattern recognition algorithms to detect lung disease or distinguish abnormal lung sounds from normal ones, reducing human diagnostic errors [4]. While this analysis may be the best alternative to traditional methods such as using a stethoscope, some challenges need to be addressed. One of the main challenges of this method is the interference that occurs when recording lung sounds. These interferences occur because pulmonary sound signals are usually recorded in hospitals and clinics and often not in acoustic environments. As a result, recorded signals are contaminated with different noises. Thus, it is necessary to reduce or eliminate any noise and disturbing environmental signals.

Distortions in the signal usually have two external and internal sources. Distortions caused by external sources, such as speech, the noise of other recording devices, a fan noise, sensor displacement, and suchlike, and their internal sources, including muscle contraction and expansion, respiratory and gastrointestinal sounds, cough, and suchlike, confuse the physician to examine the pulmonary sound signal characteristics to diagnose pulmonary disorders [5], [6]. In the noise reduction process, it is crucial to preserve unknown adventitious components that provide the most information diagnostic value during lung pathology [7], [8].

In computer analysis, environmental noises are usually considered equivalent to colored noises. When recording the LS signal, the environmental noise can be considered equivalent to white and pink noises. Gaussian white noise has constant energy at all frequencies. In mathematical terms, the Power Spectrum Density (PSD) of this noise can be represented by **Eq.1**:

$$S(f) = \frac{C_w}{|f|^0} \qquad (1)$$

where $C_w$ is a constant value, and $f$ is the frequency.

This white noise definition is ideal, but in the real world, most noises are pink noise, which has a cut-off frequency unlike white noise having the same amount of energy at all frequencies. This means that the amount of energy decreases with increasing frequency. In mathematics, we can show the PSD of pink noise with **Eq.2**:

$$S(f) = \frac{C_f}{|f|^\alpha} \qquad (2)$$

where $C_f$ is a constant value, $f$ is the frequency, and $0 < \alpha < 2$. If $\alpha = 0$, it is equivalent to white noise, and if $\alpha = 2$, it is equivalent to red noise [6].

Reviewing previous studies shows that removing white and pink noises from vital signals has been the focus of many studies. In addition, most of these studies have not considered that LS signals are noisy in nature and can be affected in the denoising process. This suggests that LS signal denoising methods need to be improved.

Over the past decade, many advances have been made in biomedical signal processing. One of the new methods in this field is Discrete Wavelet Transform (DWT), whose important multi-resolution property allows the analysis of different signals at multiple resolution levels. Extensive research has shown that DWT has many advantages in various fields, including denoising and signal compression [6], [9]–[11]. Nowadays, common wavelet threshold approaches are widely used for denoising biomedical signals such as Electrocardiography (ECG) [12]–[15] and Electromyography (EMG) [16], [17]. In addition, this method also plays a crucial role in LS signal denoising [7], [8], [18], [19]. Although many researchers have suggested DWT-based methods for LS denoising, this method also has its drawbacks. Most of these drawbacks are related to the parameters that need to be determined in advance and experimentally based on the input signal characteristics, such as the mother wavelet, decomposition levels, and the threshold value. Therefore, it is better to use alternative data-driven methods to solve the problem of parameter setting. To solve DWT problems in threshold-based denoising methods, the EMD algorithm has been proposed as a more suitable alternative [20]–[22].

Denoising of the LS signal can face various challenges, one of which is that different signals that differ in their characteristics can be recorded depending on the location of the LS signal. In addition, the performance of the denoising process is directly related to the threshold value setting due to the LS noisy nature. If this value is not selected optimally, the useful content of the LS signal may be removed in addition to noise. For this reason, denoising of the LS signal using the thresholding cannot perform well because a specific threshold value cannot be considered for all conditions. For this reason, it is necessary to use a method that does not need to set the threshold [23], [24]. To solve this problem, M. F. Pouyani et al. proposed a method based on Discrete Wavelet Transform and Artificial Neural Network (DWT-ANN) to perform denoising adaptively without specifying predefined parameters. In this method, the multi-resolution property of DWT with adaptive learning and nonlinear mapping of the ANN are used simultaneously [25].

The neural network acts as a nonlinear adaptive filter and maps the input noisy signal wavelet coefficients to the desired signal samples by adjusting its weights. As a result, the noise reduction process is performed without thresholding. This paper attempts to perform comparative filtering of LS signals using EMD and ANN. In the proposed method, the EMD helps extract the IMFs of the LS signal and give it as the input to the neural network, and then the IMFs are mapped to the clean signal samples by the neural network. The advantage of EMD over wavelet transform is that EMD acts as data-driven, and the number of extracted IMFs varies depending on the nature of the input signal and the recording location, while in wavelet transform, the number of levels is fixed and must be determined in advance. Given that some IMFs contain useful signal information and others contain signal and noise, choosing the right number of these functions is important. The neural network automatically determines the weights of each input IMF, thus, there is no need to predetermine the optimal IMFs (for example, IMFs with less noise and more information). Another advantage of the neural network is that the output of this network is a time-domain denoised signal. Unlike previous studies, to evaluate the performance of the proposed method in this study, both white and pink noises with SNRs of 0, 5, 10, 15, and 20dB were added to the clean signal, and the results were obtained compared with those of the EMD-Custom method.

The structure of this article is as follows. The second section explains the database used in this research. The third section analyzes normal lung sounds and the method proposed in this study. The fourth section deals with preprocessing and evaluation parameters. Then, the results of the implementation of the algorithms are presented and discussed in the fifth section. In the end, the sixth section represents the discussion, conclusions, and suggestions for future research.

## 2. Dataset

In this study, LS signals were recorded from five healthy individuals (five women between the ages of 25 and 35) when they were asked to sit on a bed in a silent room. These signals are recorded asynchronously from 16 different parts of the back of their chest by an electret microphone. Each recording lasted 80 seconds, including several respiratory cycles. The location diagram of this microphone can be seen in **Fig.1**. All signals are stored by an analog-to-digital converter with a 16-bit resolution and a sampling rate of 8000 Hz. Given the location and condition of the data recorded in a silent room and the fact that the individual was asked to breathe slowly to reduce additional respiratory noise, all collected data can be considered a clean signal.

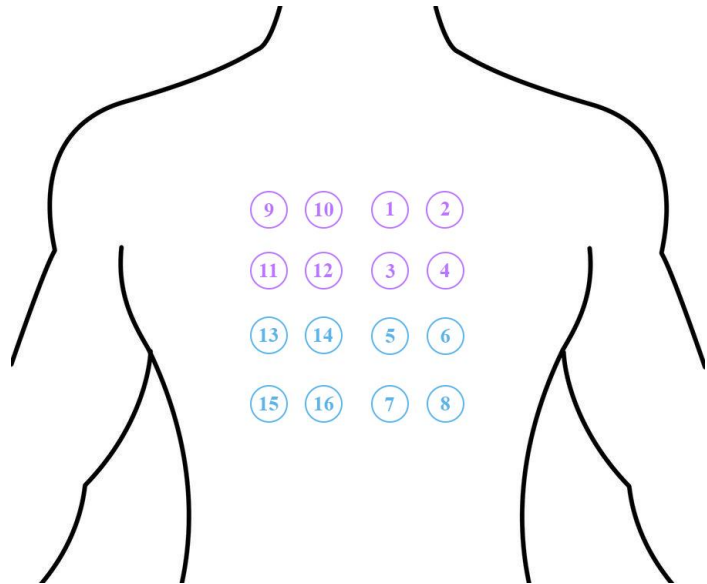

**Fig. 1:** Overview of microphone locations

**Note:** Channels colored purple and blue contain bronchovesicular and vesicular sound signals, respectively.

## 3. Methodology

### 3.1. Normal respiratory sounds

The production of respiratory sounds is related to the turbulent airflow in the respiratory tract, and these sounds can be recorded from the chest and trachea or from behind the chest [26]. Respiratory sounds can be divided into two categories: normal and abnormal breathing sounds. Turbulent airflow in the airways and changes in this flow produces normal breathing sounds, divided into four categories: tracheal, bronchial, bronchovesicular, and vesicular sounds.

The trachea sound is heard from the slit above the sternum (at the bottom of the neck) or around the neck. Vesicular sound is heard throughout the lungs, except for the sternum and between the shoulders. Bronchial sound is heard on the trachea and main bronchi, and bronchovesicular sound is heard between the bronchial and vesicular sound, on either side of the sternum and between the scapulae [27], [28].

**Fig.2** shows an overview of the location of normal lung sounds. In addition, the channels colored purple and blue contain bronchovesicular and vesicular sounds, respectively (**Fig. 1**). There are two possible ways to record a lung sound signal. The first is to record these signals from the front of the chest. The advantage of this is that it provides specific information about the lung condition, but at the same time, the heart sound interferes with the lung sound signals, making it very difficult to distinguish the primary signal. The second reason is that the heterogeneity of the chest wall varies the amplitude of the respiratory sounds at the level of the chest. For example, on the chest surface, breathing sounds cannot be well received from the bones (respiratory sounds are poorly transmitted in areas of the chest surface located on the bones). As a result, lung sounds should not be recorded from the chest bones [29], [30]. The second case is to record this signal from the back of the chest. The advantage of this mode is that the heart noise is less than the lungs noise, which will make the signal processing less challenging.

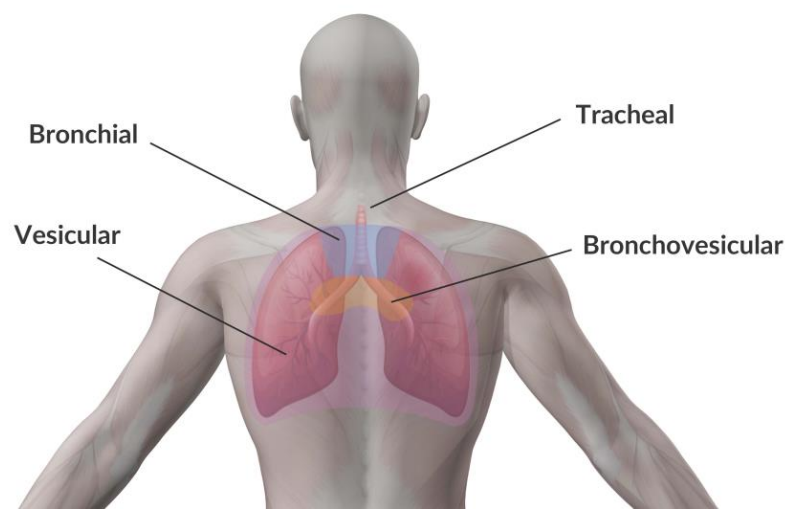

**Fig. 2:** Places to auscultate normal respiratory sounds

## 3.2. Empirical Mode Decomposition (EMD) Algorithm

Since 1822, when Fourier first stated that any nonperiodic function could be defined using weighted integrals as a set of sine and cosine functions, more complete and different methods proposed by various individuals have tried to solve the problems of their previous methods to some extent. For example, one of the main problems with the Fourier transform is that the integral relation is applied to the whole signal and the average value of a frequency is obtained in the whole signal. In fact, the Fourier transform destroys the spatial information of frequencies. To solve this problem, the Short-Time Fourier Transform (STFT) method was proposed to split the main signal into windows and convert a time-domain signal into a two-dimensional time-frequency display by applying a Fourier transform [19].

Given that determining the appropriate window length was one of the challenges of the STFT, and while the length of this window is constant during application to the signal, the Wavelet Transform (WT) was proposed to use a window with variable lengths, overcoming the problem of preset resolution. Wavelet transforms for non-static signals will be suitable with the content of unknown frequency (what frequency occurs at what moment). Several frequency intervals are obtained by applying wavelet transform on the signal depending on the number of selected decomposition levels[31].

In 1996 [32], the EMD as the most suitable method for the time-frequency analysis of nonlinear and non-stationary data was first proposed by Huang et al. As a data-driven tool, the EMD, aims the comparative expression of signals as a set of zero-mean oscillating components, called Intrinsic Mode Functions (IMF), using a sifting process. The IMFs and the residue sum achieve the signal reconstruction process. A function can be IMF if it meets two conditions [22]:

- The number of extremes equals zero points or varies by a maximum of one.
- Its integral is zero in the defined time interval.
- The method of signal analysis into the intrinsic mode functions, called the screening process, is described in the following order[33]:

1. Find the maxima and minima of $x(t)$ and use cubic spline interpolation to create the upper and lower envelopes, respectively.
2. Calculate the average envelope $m(t)$ by averaging the upper and lower envelopes.
3. Calculate the temporary local oscillation:

$$h(t) = x(t) - m(t) \qquad (3)$$

4. Calculate the average of $h(t)$. If the mean of $h(t)$ is close to zero, then $h(t)$ is considered the first IMF named $c_i(t)$. Otherwise, repeat steps (1)-(3) while using $h(t)$ for $x(t)$.
5. Calculate the residue $r(t) = x(t) - c_i(t)$
6. Repeat steps from (1) to (5) using $r(t)$ for $x(t)$ to obtain the next IMF and residue.

The decomposition process ends when the residue $r(t)$ becomes a monotonic function or constant and no longer meets the IMF's conditions.

$$x(t) = \sum_{i=1}^{N} h_n(t) + r_N(t) \quad (4)$$

where $N$ is the number of IMFs, $h_n(t)$ is IMF, and $r_N(t)$ is the last residue.

The sifting process is continued until the last residual is either a monotonic function or a constant.

### 3.3. EMD Based Noise Reduction

WT is widely used in physiological signal-denoising methods and has a good ability for noise reduction if there is spectral overlap between signal and noise. In addition to its advantages, however, there are some disadvantages related to the parameters that need to be determined experimentally in advance based on the input signal characteristics, such as the mother wavelet, the number of level decompositions, and the threshold value regulation. Therefore, it is better to use alternative data-driven methods to solve the problem of parameter settings. One of the best alternatives to the WT-based method is the EMD-based denoising method, the main advantage of which is that the base functions are derived from the input signal, unlike the wavelet transform method, which is fixed. Recently, the EMD method has been used for signal denoising in many applications such as biomedical and acoustic signals. Noise-related components in the noisy signal are often focused on the first IMFs (high-frequency IMFs), and the useful signal information is often focused on the last IMFs (low-frequency IMFs). As a result, the noise reduction method can be based on the partial structure of the signal using only the last relevant IMFs [21], [34]. The analysis of the noisy LS signal to the IMFs is shown in **Fig. 3**.

### 3.3.1. EMD-Soft thresholding and EMD-Hard Thresholding

One of the most widely used EMD-based denoising algorithms is the threshold denoising algorithm, in which a noise-contaminated signal with a finite length is defined as:

$$y(t) = x(t) + \eta(t) \quad (5)$$

where $x(t)$ is the desired signal and $\eta(t)$ is the white Gaussian noise.

In the EMD-Soft thresholding method, the noisy LS signal $y(t)$ was first decomposed into noisy IMFs $c_{ni}(t)$. These noisy IMFs were thresholded by a soft or hard function to obtain an estimation of the denoised IMFs $\widehat{(c_i)}(t)$ of the desired signal. The definition of the hard and soft thresholding is given below:

A direct application of wavelet hard thresholding [35] in the EMD case:

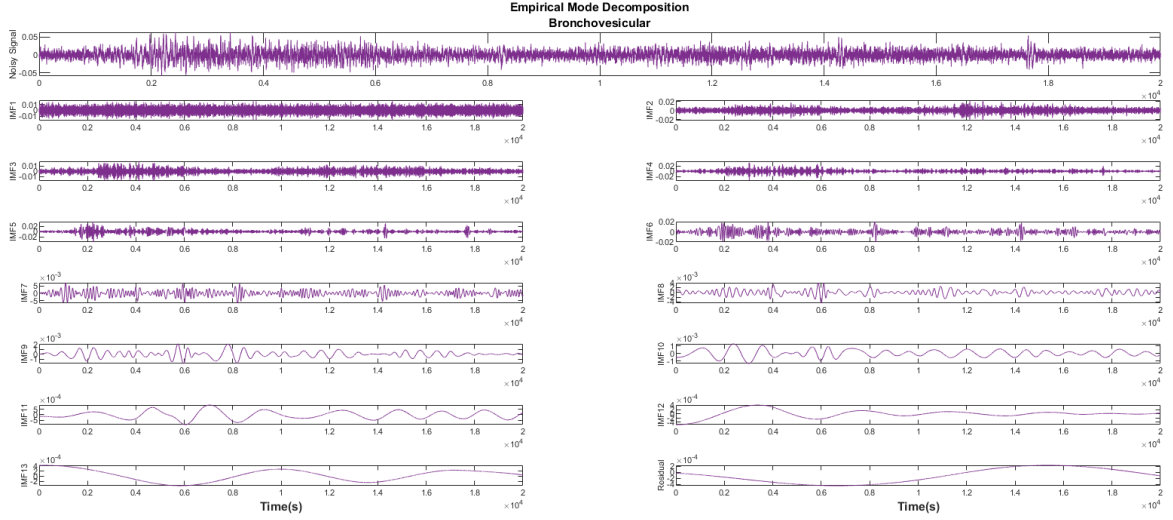

**(a)**

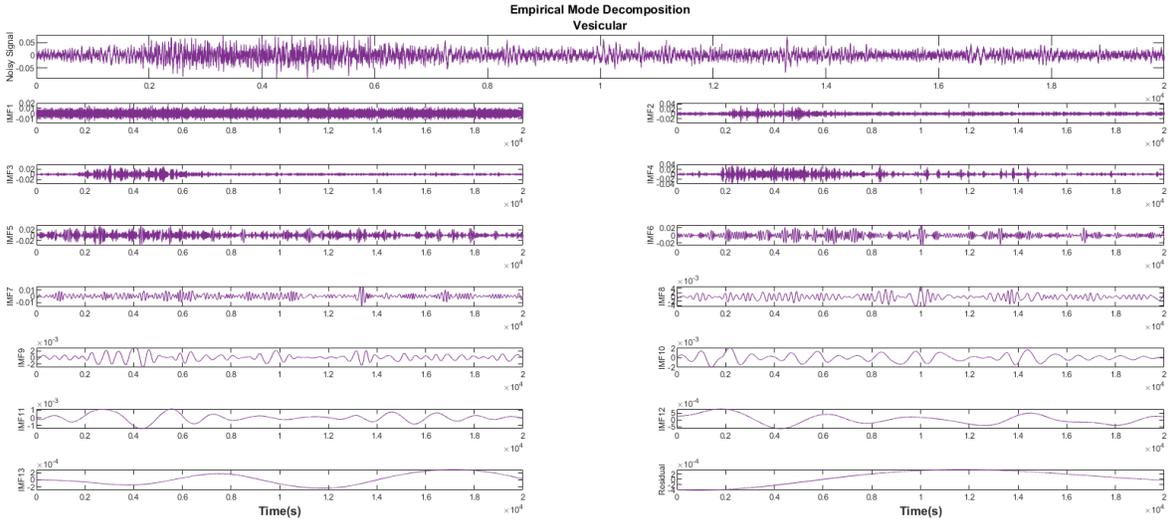

**(b)**

**Fig. 3:** Decomposition of noisy signal into IMFs using the EMD algorithm for two (a) Bronchovesicular and (b) Vesicular signals

$$\widehat{c}_i(t) = \begin{cases} c_{ni}(t) & if \ |c_{ni}(t)| > \tau_i \\ 0 & if \ |c_{ni}(t)| \leq \tau_i \end{cases} \quad (6)$$

A direct application of wavelet soft thresholding [35] in the EMD case:

$$\widehat{c}_i(t) = \begin{cases} c_{ni}(t) - \tau_i & if \ c_{ni}(t) \geq \tau_i \\ 0 & if |c_{ni}(t)| < \tau_i \\ c_{ni}(t) + \tau_i & if \ c_{ni}(t) \leq -\tau_i \end{cases} \quad (7)$$

According to the hard thresholding definition, all IMFs smaller than the threshold are removed, and the rest of the coefficients are kept unchanged. Similarly, all IMFs that are smaller than the threshold are removed in soft thresholding, and the remaining IMFs are scaled according to **Eq. 7**.

The threshold value should be selected so that it effectively retrieves the content of the original signal in addition to removing noise from the noisy signal. If the threshold value is very high, it will delete the main contents of the signal, and if the threshold value is very low, denoising will not work correctly[36]. Next, a discussion is presented on the universal threshold method proposed for selecting the optimal threshold value [37].

$$\tau_i = C\sqrt{E_i 2\ln(n)} \qquad (8)$$

where $C$ is a constant depending on the type of signal that was set to 0.7 in this work, $n$ is the length of the signal and $E_i$ is given by:

$$\hat{E}_i = \frac{E_1^2}{0.719} 2.01^{-i} \quad, i = 2, 3, 4 \cdots N \qquad (9)$$

where $E_1^2$ is the energy of the first IMF defined by:

$$E_1^2 = \left(\frac{median(|c_{n1}(t)|)}{0.6745}\right)^2 \qquad (10)$$

A reconstruction of the denoised signal is given by:

$$\hat{x}(t) = \sum_{i=1}^{N} \hat{c}_i(t) + r_N(t) \qquad (11)$$

### 3.3.2. EMD-Custom Thresholding

One of the main problems with hard (or soft) thresholding is time-frequency discontinuities, leading to the production of annoying artifacts and further degradation of the output signal. A modified custom thresholding function is provided to solve this problem. It can be defined a modified custom thresholding function as follows [37]:

$$\hat{c}_i(t) = \begin{cases} c_{ni}(t) - sgn(c_{ni}(t))[1-\alpha]\tau_i & if\ |c_{ni}(t)| \geq \tau_i \\ 0 & if\ |c_{ni}(t)| \leq \gamma \end{cases} \qquad (12)$$

where: $0 < \gamma < \tau_i$ and $0 \leq \alpha \leq 1$.

**Fig. 4** shows an overview of the EMD-Custom threshold-based denoising method.

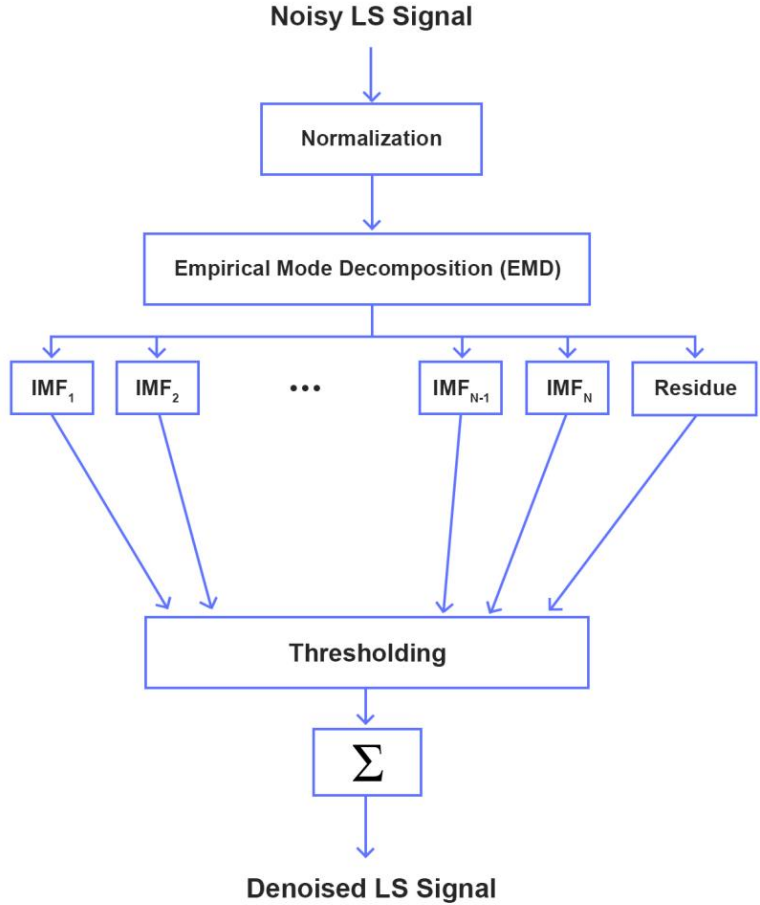

**Fig. 4:** Overview of the denoising method based on EMD-Custom thresholding

### 3.4. EMD-ANN Noise Reduction Method

As mentioned in the previous sections, one of the most widely used methods for LS signal noise reduction is the threshold method based on EMD. However, this method has some problems, such as mode mixing, in which one of the following conditions may occur [38]:

1. Formation of two or more dissimilar oscillating components in an IMF.

2. The formation of two or more similar oscillating components in several IMFs.

The phenomenon of mode mixing reduces the efficiency of the EMD decomposition process; in other words, none of the IMFs are pure (containing only useful signal information). On the other hand, one of the main factors influencing the quality of signal denoising is thresholding and selecting the appropriate thresholding function. Because lung sound is noisy in nature, and the existence of several types of lung sounds with different characteristics leads to different IMFs, the thresholding method cannot separate them and perform noise removal well. In fact, the IMFs produced by noisy bronchovesicular and vesicular signals are different as shown in **Fig. 5**. It is impossible to consider a specific pattern for both of them, using it for selecting threshold values.

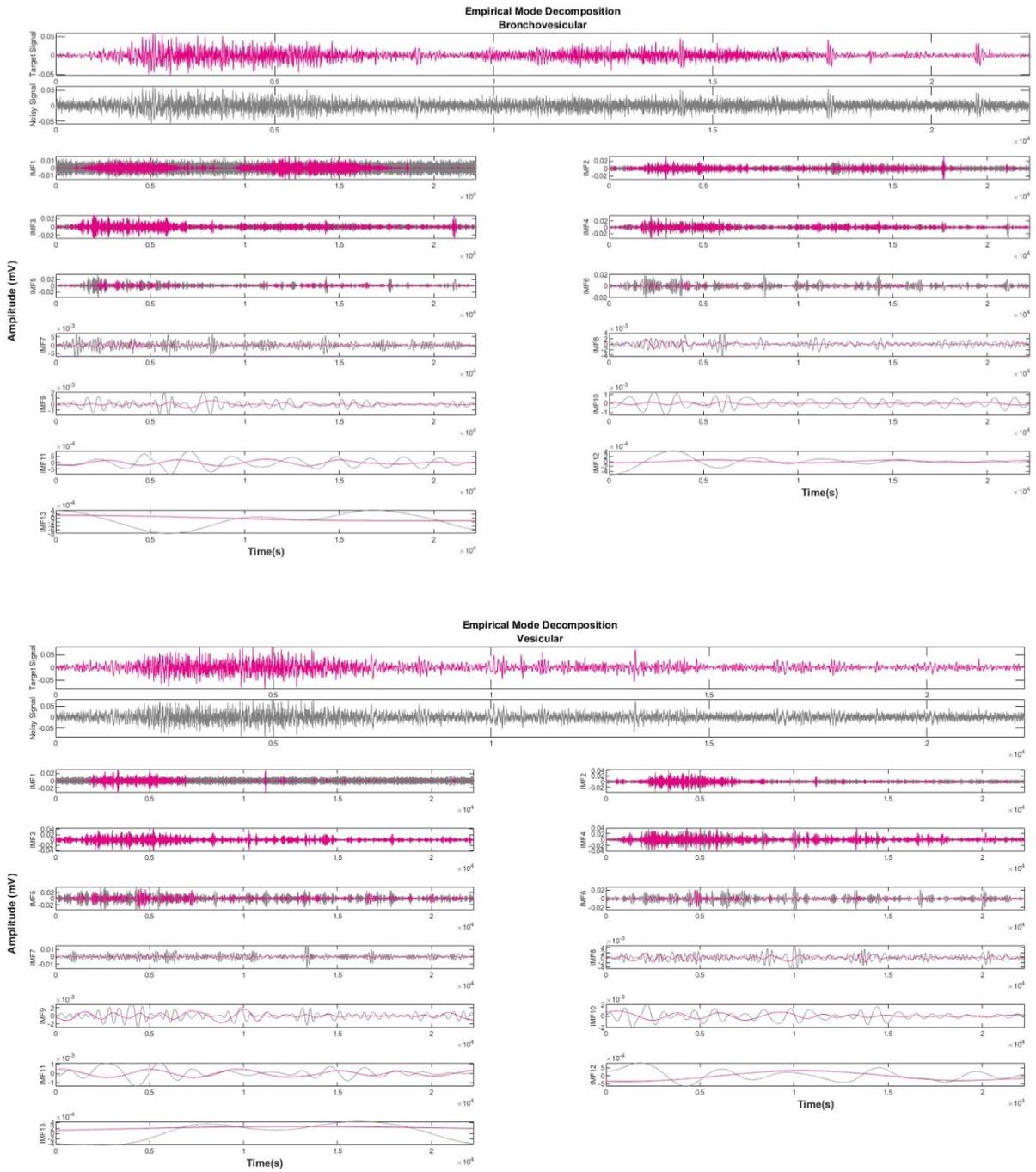

**Fig. 5:** Decomposition of the noisy signal and the target (desired or clean) signal to IMFs and mapping them to 13.

For example, it may be considered a threshold for vesicular sound IMF1; if the same threshold is used for the bronchovesicular signal IMF1, it may be considered part of the expiratory sound as noise and remove its main content. The idea of using Empirical Mode Decomposition and Artificial Neural Networks (EMD-ANN) is proposed to solve these problems.

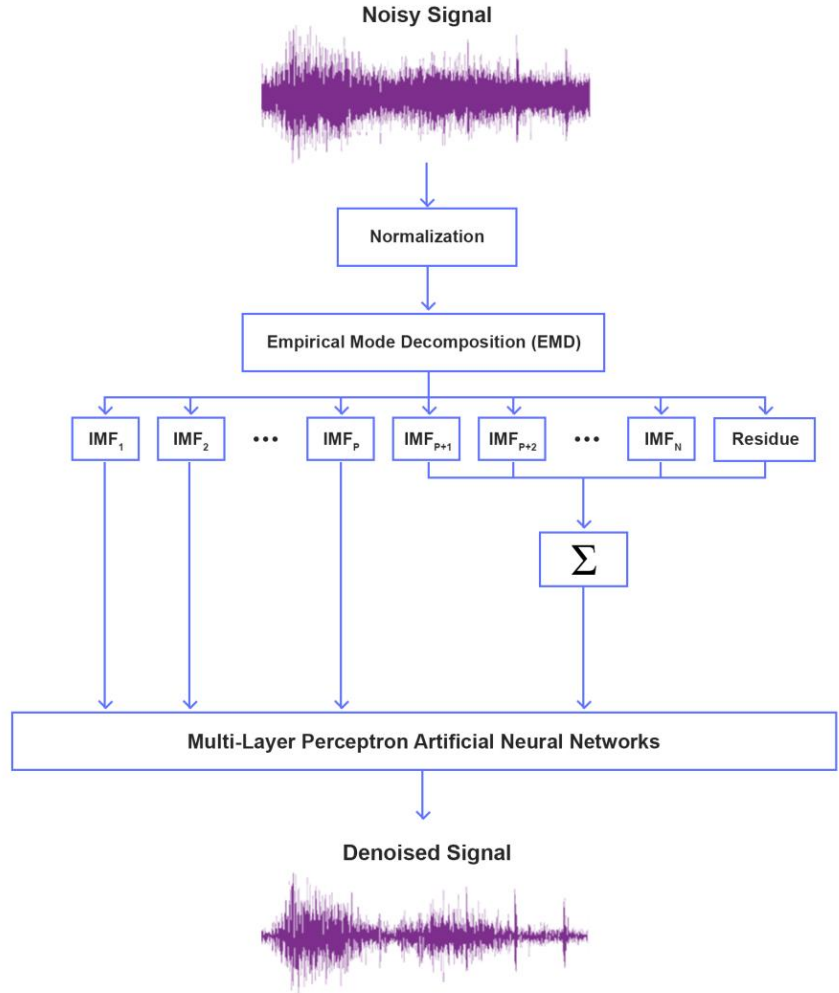

**Fig. 6:** Overview of the proposed EMD-ANN noise reduction method, p = 12.

The neural network tries to obtain the main content of lung sounds from the noisy IMF. In the training phase, therefore, the input to the neural network is the IMF associated with each sample of the noisy signal, and its output is the corresponding sample from the clean signal. Thus, during the test phase, the NN learns to remove noise from the noisy signal sample on a sample-by-sample basis. According to an overview of the proposed method in **Fig. 6**, it is first necessary to give the preprocessed signal as input to the EMD to obtain the signal IMFs, which are now given as the input to the MLP neural network.

After the IMFs enter the neural network, it adjusts their weights to reduce the Mean Squared Error (MSE) between the network and the desired output by considering these IMFs as features and comparing them with the target, which is the desired signal. Because the output layer activation function is linear, all outputs of the last hidden layer are summed and considered the network output with no changes. For this reason, the neural network output is the time-domain denoised signal.

## 4. Experiment

As mentioned earlier, 80 seconds of LS signals were recorded from 16 different areas of each participant in this study. Because the recorded signals include several respiratory cycles, one cycle is randomly selected from each person. Twelve out of the 16 chosen cycles from each person are considered as training data and the rest as test data. Next, between 13 and 15 IMFs are obtained by applying EMD on each respiratory cycle. Because the number of IMFs is different for each respiratory cycle, all IMFs have been reduced to 13 to avoid errors in the calculations and facilitate the generalizability of this method. (For this purpose, when the number of IMFs is more than 13, it is necessary to sum the IMFs 13 and later make it equal to the 13th IMF). Now, all the IMFs obtained from the training data are concatenated together and given as the input to the neural network. In the test phase, the IMFs are each time extracted from one of the respiratory cycles so that they can be given as the input to the neural network after reducing their number to 13. The average results obtained from examining test signals are announced as the final result.

### 4.1. Data Preprocessing

In the preprocessing phase, because the useful content of the LS signal is only in the range of 20 to 2000 Hz, the sampling rate can be reduced from 8000 to 4000 Hz to reduce the computational volume [21]. The spectrograms of clean bronchovesicular and vesicular signals can be seen in **Fig.7**. Then, Gaussian white and pink noises with SNRs = 0, 5, 10, 15, and 20dB were added to these data as these two noises added to the signals are the best simulation of the noises in the environment where the LS signal is recorded [6]. Then, it is necessary to normalize the data to increase the learning speed of the neural network. The normalized signal amplitude will be in the range [-1,1] with this normalization using the following formula [24]:

$$x_{norm}(t) = 2\left(\frac{x(t) - x_{min}}{x_{max} - x_{min}}\right) - 1 \qquad (13)$$

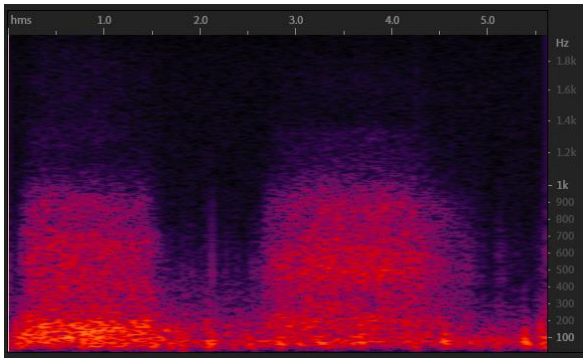   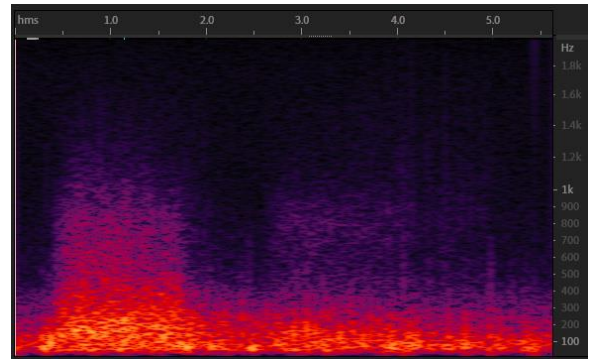

(a) Spectrogram of the clean bronchovesicular signal      (b) Spectrogram of the clean vesicular signal

**Fig.7:** Clean bronchovesicular and vesicular signal spectrograms

where $x(t)$ is the noisy LS signal, $x_{norm}(t)$ is the normalized signal, $x_{min}$ and $x_{max}$ are respectively the minimum and maximum values of the noisy LS signal, and $t$ is the index of the signal samples. From now on, $x_{norm}(t)$ is used instead of $x$ in the following steps.

## 4.2. System Evaluation

As with many other papers, the SNR and Fit parameters are used in this study to evaluate the performance of the proposed method [23]. The SNR is a parameter that can compare the energy level and the noise of the desired signal so that an increase in SNR indicates low noise in the signal. In other words, there is a negative correlation between the SNR value and the amount of noise in the signal. The SNR formula is as follows:

$$SNR(x_d, y) = 10 log_{10} \frac{\sum_i^N x_d(t)^2}{\sum_i^N (y(t) - x_d(t))^2} \tag{14}$$

where $N$ is the number of signal samples, $x_d$ is the desired signal, and y is the denoised signal.

The Fit parameter indicates what part of the original content remains so that 0% and 100% indicate the loss and preservation of all the main LS content, respectively. The Fit formula is as follows:

$$Fit = 100 \times (1 - \frac{\sum_i^N (y(t) - x_d(t))^2}{\sum_i^N (x_d(t) - \frac{1}{N}\sum_i^N x_d(t))^2}) \tag{15}$$

## 5. Result

### 5.1. Select the appropriate neural network structure

In individual models, the network is trained with a specific SNR and then evaluated with the same SNR. In other words, since Gaussian white noise with five different SNRs has been added to the clean signal, there are five independent individual models. The advantage of using the individual model is that the network is well trained with the input noise and, as a result, will have a good ability to denoise the signal. To design an individual model, it is first necessary to determine the structure of the neural network. Hence, signals infected with Gaussian white noise were used to determine the network structure.

According to **Table 1**, nine proposed structures are considered among which ANN1, ANN2, and ANN3 networks have a hidden layer, and ANN4, ..., ANN9 networks have two hidden layers. As mentioned in the description of the proposed method, the number of IMFs after mapping is equal to 13. As a result, the number of input layer neurons is 13 due to the lack of bias. On the other hand, since this is a regression problem, the network output has only one neuron, and its activation function is linear. The Levenberg-Marquardt algorithm is used as the backpropagation to determine the neural network weights. The gradient descent technique is also used to reduce the squares of the cost function error. It should be noted that the number of repetitions of network training is equal to 200. The hyperbolic tangent sigmoid transfer function was used as the hidden layer activation function, and the linear transfer function was used as the output layer activation function.

Table 1: Definition of Neural Networks structures

|       | Input Layer | First Hidden Layer | Second Hidden Layer | Output Layer |
|-------|-------------|--------------------|---------------------|--------------|
| ANN 1 | 13          | 35                 | -                   | 1            |
| ANN 2 | 13          | 65                 | -                   | 1            |
| ANN 3 | 13          | 95                 | -                   | 1            |
| ANN 4 | 13          | 25                 | 15                  | 1            |
| **ANN 5** | **13**  | **25**             | **20**              | **1**        |
| ANN 6 | 13          | 25                 | 25                  | 1            |
| ANN 7 | 13          | 35                 | 15                  | 1            |
| ANN 8 | 13          | 35                 | 20                  | 1            |
| ANN 9 | 13          | 45                 | 10                  | 1            |

Table 2: Results obtained from the examination of different neural network structures using individual models with white noise

| NN Structure | SNR = 5 | | SNR = 10 | | SNR = 15 | |
|--------------|---------|---------|----------|---------|----------|---------|
|              | SNR (dB) | Fit (%) | SNR (dB) | Fit (%) | SNR (dB) | Fit (%) |
| ANN 1 | 13.54 | 95.05 | 16.83 | 97.68 | 20.32 | 98.95 |
| ANN 2 | 13.64 | 95.15 | 17.21 | 97.86 | 20.41 | 98.98 |
| ANN 3 | 13.54 | 95.01 | 17.19 | 97.85 | 20.74 | 99.04 |
| ANN 4 | 13.69 | 95.20 | 17.49 | 97.99 | 21.04 | 99.12 |
| **ANN 5** | **13.80** | **95.32** | **17.64** | **98.04** | **21.53** | **99.20** |
| ANN 6 | 13.77 | 95.28 | 17.49 | 97.99 | 21.24 | 99.15 |
| ANN 7 | 13.72 | 95.24 | 17.52 | 98.00 | 21.22 | 99.14 |
| ANN 8 | 13.75 | 95.26 | 17.60 | 98.02 | 21.45 | 99.18 |
| ANN 9 | 13.67 | 95.19 | 17.51 | 97.99 | 21.16 | 99.13 |

According to **Table 2**, it is clear that networks with two hidden layers perform better. By examining more closely and comparing the results obtained from the performance of networks with two hidden layers, it is clear that the performance of the ANN5 network having 25 and 20 neurons in the first and the second layers, respectively, has better results than the other networks. As a result, it is selected as the optimal network.

## 5.2. Noise Reduction using Combined Model

As mentioned in the introduction, some of the noise in the recording room, such as noise from other devices, a fan, speech, and suchlike can be prevented. Since there is internal noise and the recording room is not entirely acoustic, there are unavoidable noises that can be considered Gaussian white and pink noises[18], [25], [37], [39]. One of the main problems that these noises can cause is that they combine with the LS signal and make it difficult and challenging to distinguish the main content of the LS from these signals. On the other hand, the amount of noise in the environment can vary from place to place and from time to time. Therefore, the performance of a network trained with a specific SNR cannot have a real application. For this reason, it is necessary to replace the individual model with other models, which are introduced in this section.

In the training process of the combined model, the proposed neural network is trained simultaneously with different SNRs. More precisely, the network input SNR in different cycles includes one of the values 0, 5, 10, 15, and 20dB. The trained network will now be able to reduce the input signals noise with SNRs of 0, 5, 10, 15, and 20dB.

**Table 3** shows the results obtained from deionizing the LS signal infected with white and pink noises using individual and combined models. In each table row, the SNR equals the value used for training and testing the network. However, the neural network was trained with 0, 5, 10, 15, and 20dB SNRs in the combined model and tested with values written in each row. The results of this table show the ability of both models to reduce noise. In addition, a closer look at **Fig. 8** reveals that at low SNRs that indicate more noise in the signal, the performance of both models is better in different noises, and both models were able to increase the SNR to 10.22 and 9.41dB for white noise and 8.74 and 8.23dB for pink noise, respectively, which decreases with increasing the input SNR. From the Fit parameter, it can also be concluded that both models can retain a large part of the main content of the LS signal. Thus, it can be concluded that although the combined model in different SNRs has always had less improvement than the individual models, it has left a good performance and can denoise the noisy LS signal well. From a practical point of view, this is important because there is no need to calculate the SNR for signal denoising.

**Table 3:** Results of white and pink noise reduction using individual and combined models

| Input SNR (dB) | White Noise | | | | Pink Noise | | | |
|---|---|---|---|---|---|---|---|---|
| | SNR (dB) | | Fit (%) | | SNR (dB) | | Fit (%) | |
| | IND-M | COM-M | IND-M | COM-M | IND-M | COM-M | IND-M | COM-M |
| SNR = 0 | 10.22 | 9.41 | 89.45 | 87.22 | 8.74 | 8.23 | 85.49 | 83.53 |
| SNR = 5 | 13.80 | 13.23 | 95.32 | 94.67 | 12.18 | 11.31 | 93.35 | 91.86 |
| SNR = 10 | 17.64 | 16.76 | 98.04 | 97.63 | 15.81 | 14.63 | 97.11 | 96.36 |
| SNR = 15 | 21.53 | 19.53 | 99.20 | 98.71 | 17.22 | 17.19 | 97.99 | 98.03 |
| SNR = 20 | 24.86 | 21.01 | 99.63 | 98.86 | 20.67 | 20.45 | 98.87 | 99.06 |

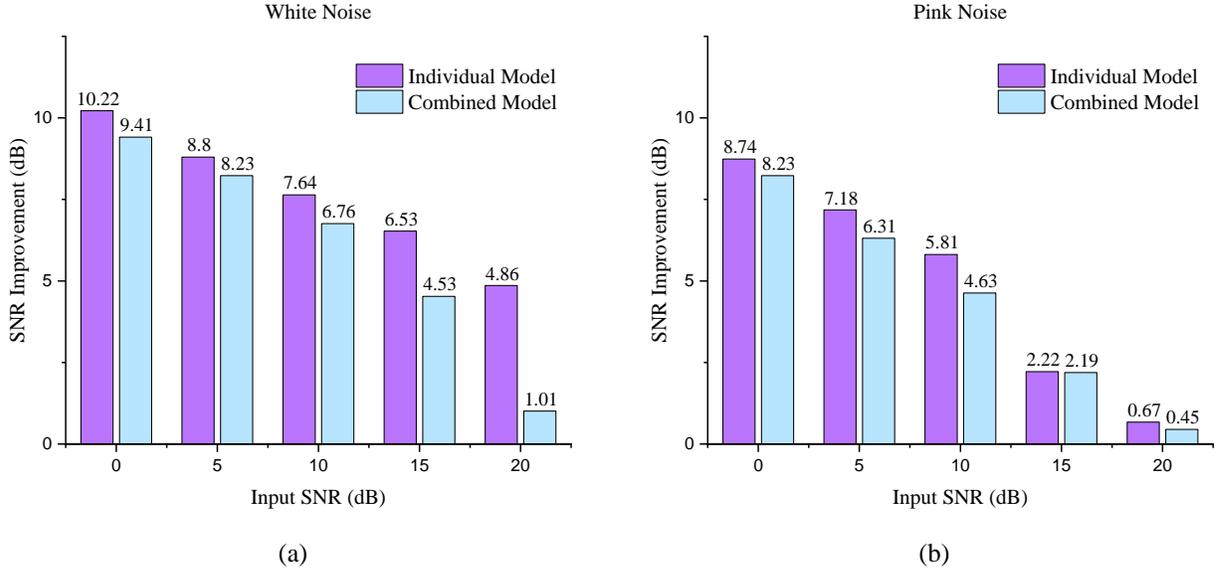

**Fig.8:** SNR improvement charts using individual and combined models

**Figs. 9**, **10**, **11**, and **12** show the bronchovesicular and vesicular sound signal spectrograms, which are infected with white and pink noises before and after the noise reduction process using the combined model. The combined model was able to retain much of the main LS content in addition to reducing noise. A comparison between the denoised signal spectrogram and the clean signal spectrogram shows that the combined model can well recover the main content in different SNRs.

According to **Figs. 10** and **12**, although the vesicular sound exhalation is much weaker than the inhalation and is very similar to noise, the output of the proposed method shows that this method reduces noise and retains the main content of the signal.

It is noteworthy that although pink noise has a non-uniform PSD, the combined model has been able to reduce both pink and white noises. However, the results are less improved than white noise at high SNRs.

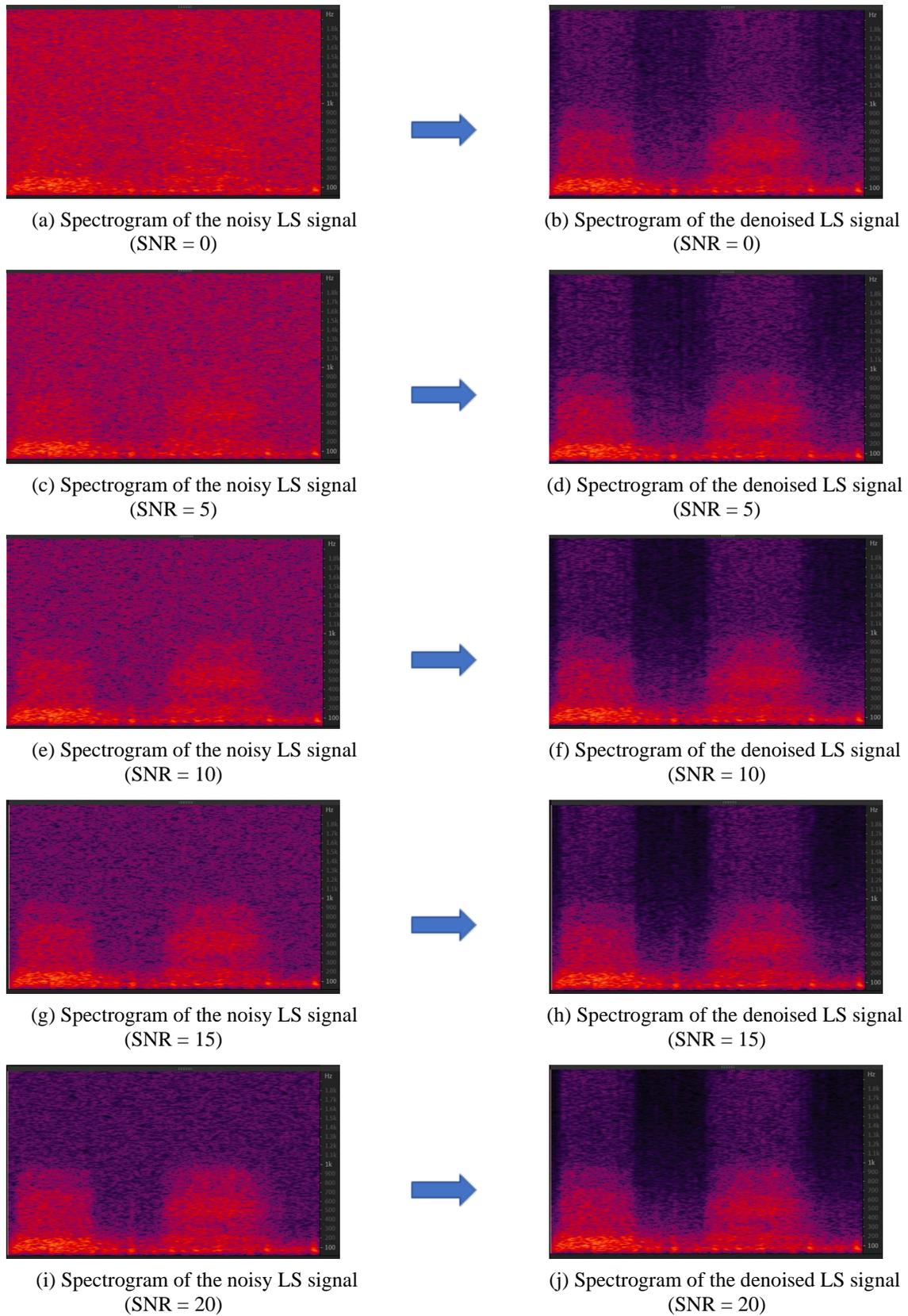

**Fig.9:** Spectrograms of white noise-infected bronchovesicular signals, before and after denoising with the combined model

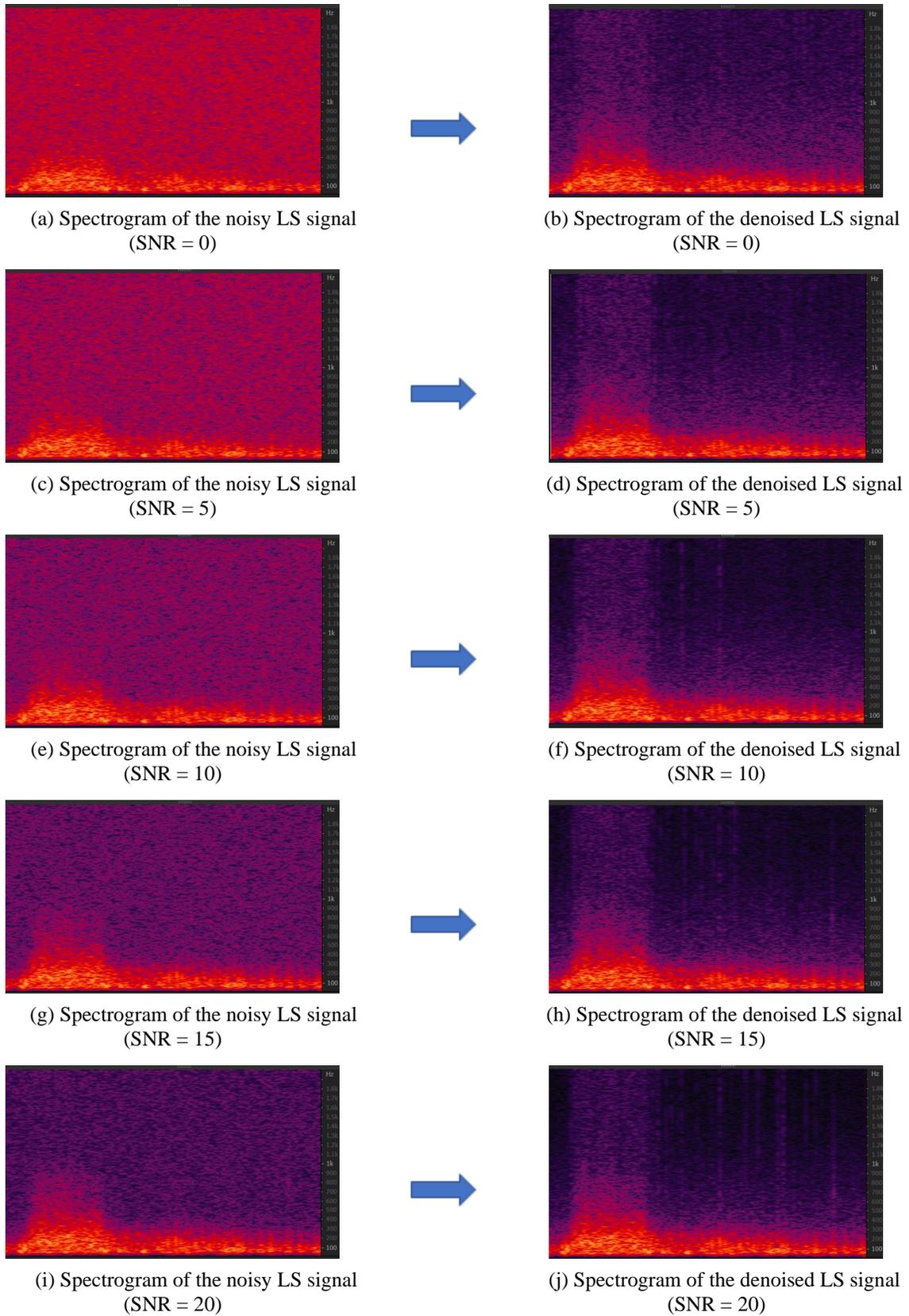

(a) Spectrogram of the noisy LS signal (SNR = 0)　　(b) Spectrogram of the denoised LS signal (SNR = 0)

(c) Spectrogram of the noisy LS signal (SNR = 5)　　(d) Spectrogram of the denoised LS signal (SNR = 5)

(e) Spectrogram of the noisy LS signal (SNR = 10)　　(f) Spectrogram of the denoised LS signal (SNR = 10)

(g) Spectrogram of the noisy LS signal (SNR = 15)　　(h) Spectrogram of the denoised LS signal (SNR = 15)

(i) Spectrogram of the noisy LS signal (SNR = 20)　　(j) Spectrogram of the denoised LS signal (SNR = 20)

**Fig.10:** Spectrograms of white noise-infected vesicular signals, before and after denoising with the combined model

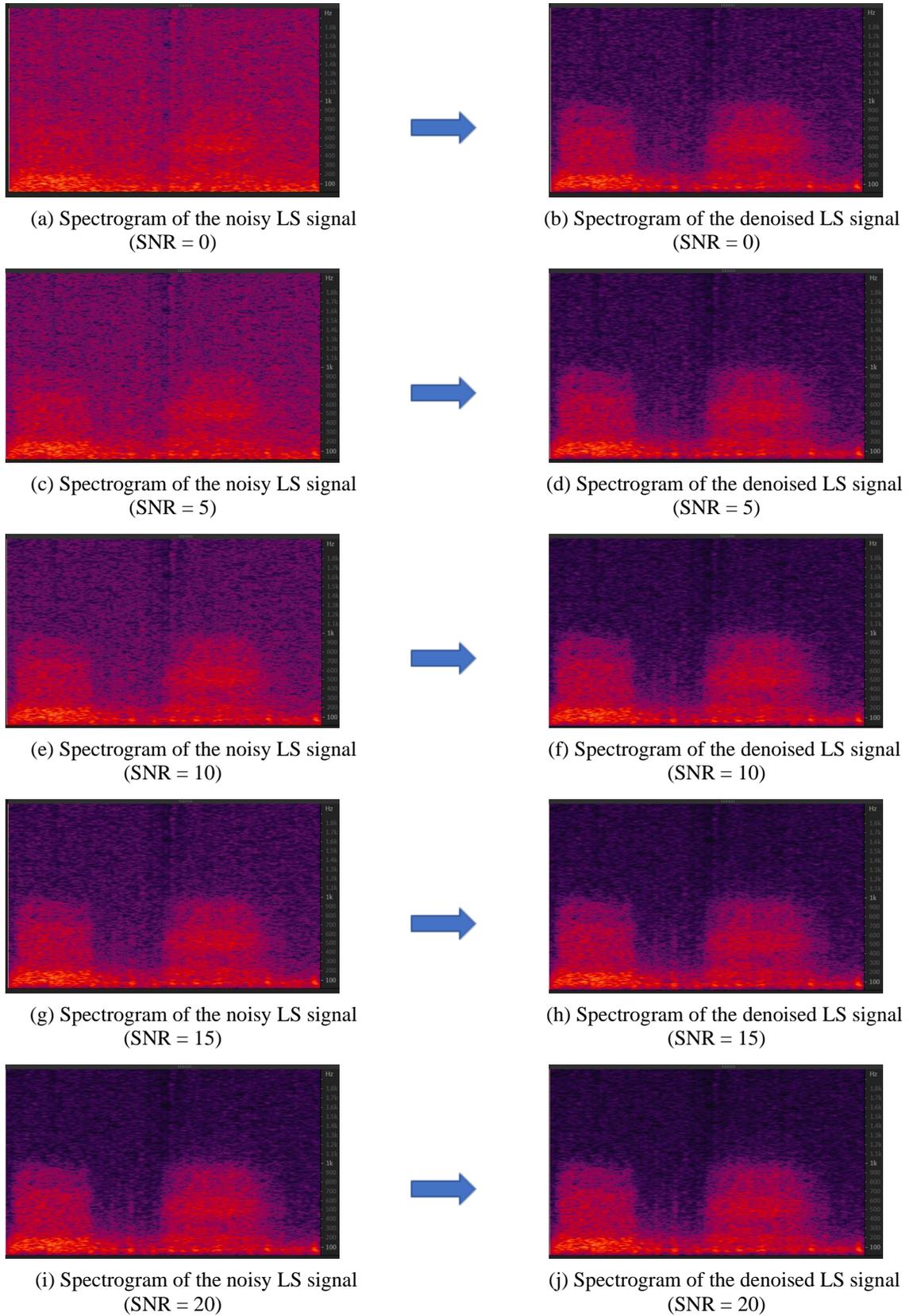

**Fig.11:** Spectrograms of pink noise-infected bronchovesicular signals, before and after denoising with the combined model

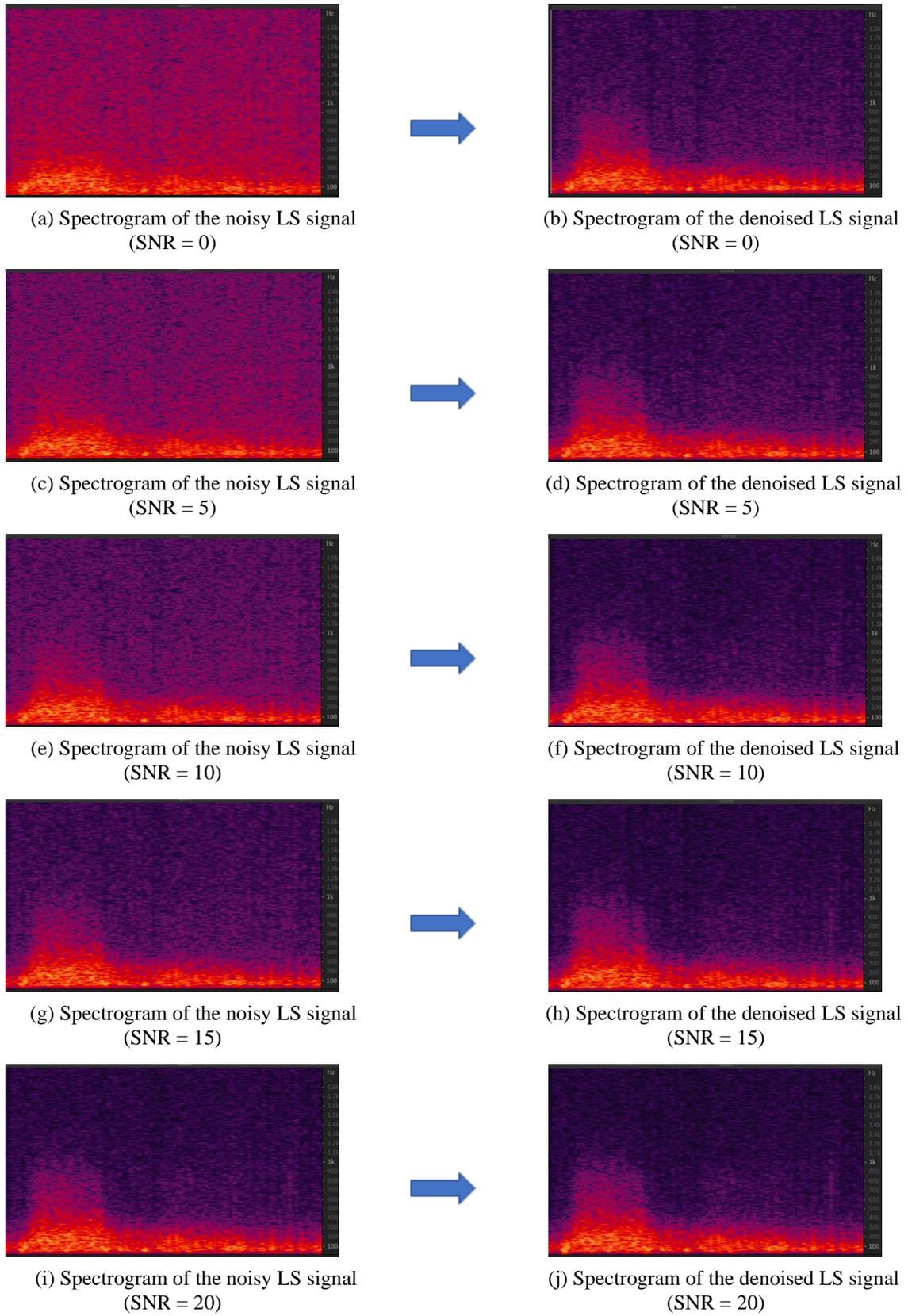

(a) Spectrogram of the noisy LS signal (SNR = 0)  (b) Spectrogram of the denoised LS signal (SNR = 0)

(c) Spectrogram of the noisy LS signal (SNR = 5)  (d) Spectrogram of the denoised LS signal (SNR = 5)

(e) Spectrogram of the noisy LS signal (SNR = 10)  (f) Spectrogram of the denoised LS signal (SNR = 10)

(g) Spectrogram of the noisy LS signal (SNR = 15)  (h) Spectrogram of the denoised LS signal (SNR = 15)

(i) Spectrogram of the noisy LS signal (SNR = 20)  (j) Spectrogram of the denoised LS signal (SNR = 20)

**Fig.12:** Spectrograms of pink noise-infected vesicular signals, before and after denoising with the combined model

## 5.3. Arbitrary noise reduction using the combined model

In the previous section, the proposed network was evaluated only with specific SNRs (such as 0, 5, 10, 15, and 20dB), and the network was tested with the same type of noise as it was trained (white or pink noise). Given that the results showed the positive performance of the proposed network, it was necessary to examine the network performance in more realistic conditions. For this purpose, different models are studied in this section, each of which has been trained with SNRs with 0, 5, 10, 15, and 20dB and then tested with signals with SNRs in the range of -2 to 20dB.

The proposed combined model is evaluated by introducing four different models: 1) white model (white) trained with white noise and tested with white noise, 2) pink model (pink) trained by pink noise and tested with pink noise, 3) white-pink model (white) trained with a combination of white and pink noises and tested with white noise, and 4) white-pink model (pink) trained with a combination of white and pink noises and tested with pink noise. The results of this evaluation can be seen in **Fig. 13**.

The examination of this diagram shows that all these models have improved the output SNR in the whole range of -2 to 20dB, which is better in low SNRs, and this improvement is minor with increasing the input SNR. It can also be seen that all models are more successful in improving signal SNRs infected with white than pink noise, which can be attributed to the uniformity of the white noise PSD, making noise removal easier for the network. The white-pink model is most similar to the real recording environment, and using this model does not need to determine the SNR or even the type of input noise. This model will also be able to denoise the signal infected with white and pink noises with different SNRs in the range of -2 to 20dB.

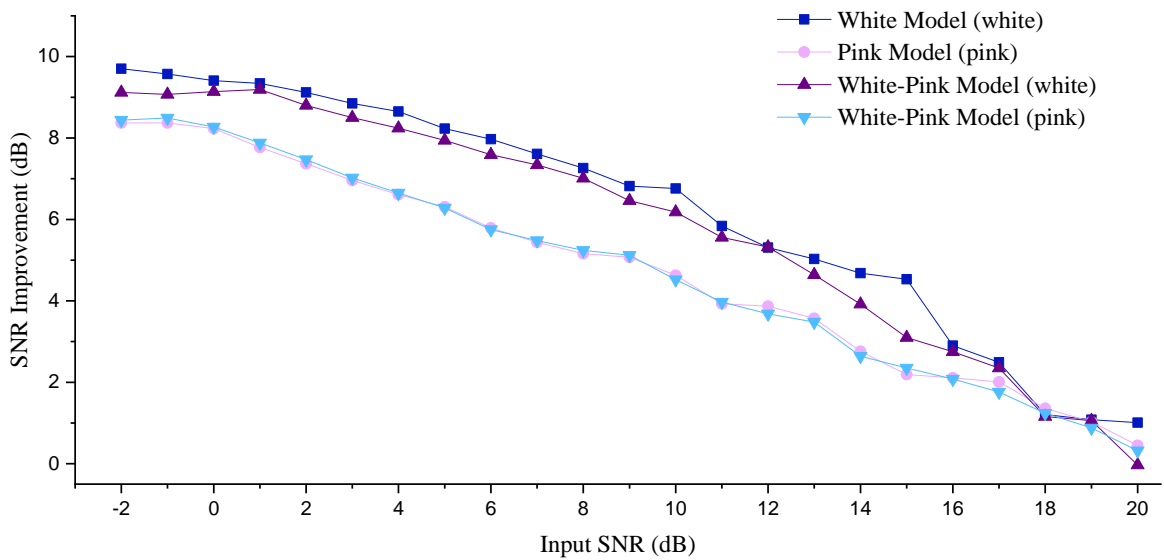

**Fig.13:** SNR improvement diagram using different combined models. The model names are based on the noise type with which the network is trained, and the name in parentheses indicates the type of noise in the test phase.

**Table 4:** Results obtained by comparing the EMD-ANN combined model with the EMD-Custom method for White and pink Noise

| Input SNR (dB) | White Model | | | | Pink Model | | | |
|---|---|---|---|---|---|---|---|---|
| | SNR (dB) | | Fit (%) | | SNR (dB) | | Fit (%) | |
| | EMD-ANN | EMD-Custom | EMD-ANN | EMD-Custom | EMD-ANN | EMD-Custom | EMD-ANN | EMD-Custom |
| SNR = 0 | 9.41 | 5.89 | 87.22 | 74.25 | 8.23 | 4.31 | 83.53 | 62.96 |
| SNR = 5 | 13.23 | 9.97 | 94.67 | 89.92 | 11.31 | 8.56 | 91.86 | 86.08 |
| SNR = 10 | 16.76 | 13.00 | 97.63 | 94.99 | 14.63 | 11.89 | 96.36 | 93.53 |
| SNR = 15 | 19.53 | 15.93 | 98.71 | 96.78 | 17.19 | 14.20 | 98.03 | 96.20 |
| SNR = 20 | 21.01 | 16.28 | 98.86 | 97.04 | 20.45 | 15.16 | 99.06 | 96.95 |

## 5.4. Comparison of EMD-ANN and EMD-Custom

This section examines the EMD-ANN combined model and the EMD-Custom method [37] with signals infected with white and pink noise.

The results in **Table 4** and **Fig. 14(a)** indicate that the proposed method can improve the output SNR at 0, 5, 10, 15, and 20 dB while the EMD-Custom method in high SNRs cannot improve the SNR value and even reduces it. As such, the output SNR is reduced by -3.72dB at SNR = 20dB because the threshold value in this method is calculated automatically using Eq. (8-10). Besides, the Fit values obtained from these two models reveal that at low SNRs such as SNR = 0dB, a large part of the main content of the LS signal is lost using the EMD-Custom method. However, the EMD-ANN has retained 87.22% of the main content.

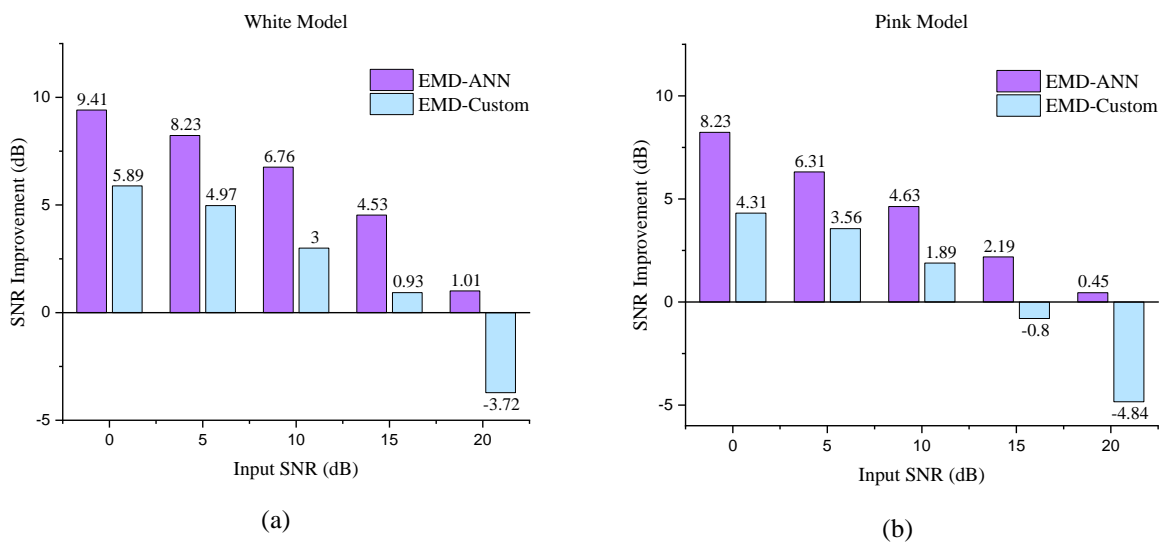

**Fig. 14:** SNR improvement diagrams using EMD-ANN combined model trained on (a) white, and (b) pink noise and comparison with EMD-Custom method

By carefully examining the second column of **Table 4** and **Fig. 14(b)**, it can be seen that the EMD-ANN performance for pink noise can improve the SNR value in all input SNRs. However, the EMD-Custom method has a poorer performance in reducing pink noise, and this model has reduced the output SNR at high SNRs. In addition, examining the results of the Fit parameter still shows that the proposed model can maintain the main content of the LS signal by 83.53% even at SNR = 0dB, while the EMD-Custom method maintained only 62.96% of the main content.

## 6. Discussion and Conclusion

The primary purpose of this paper was to design an algorithm that can denoise the LS signal infected with white or pink noises. For sound signal denoising, two thresholding methods are typically used based on WT [7], [18], [19] or EMD [21], [22]. Decomposing a signal into its orthogonal components is one of the most important steps in the noise reduction process. The quality of the denoised signal is highly dependent on the components resulting from its decomposition. An improper decomposition can lead to the removal of useful content and the retention of irrelevant information in the signal. In the WT method, the amount of useful content extracted from the signal depends on the similarity of the mother wavelet with the original signal. On the other hand, in most wavelet functions, there are parameters that are very difficult to set. EMD-based methods have been introduced to solve these problems. In EMD, signals are adaptively decomposed into their constituent modes based solely on the local properties of data without the use of orthogonal bases; however, these methods are not without problems. Considering the noisy nature of LS signals, the main challenge in these two methods is to choose the appropriate threshold for noise reduction. Although the thresholding method can reduce the noise in the LS signal, it needs to change the threshold value according to the input SNR, which makes it inefficient in real environments. In addition, since the LS signal can be recorded from different parts of the chest, which also have different characteristics, it is more challenging to determine the threshold because it is necessary to set a threshold for each location. Therefore, it is necessary to propose a method that does not need to set the threshold value.

This paper proposes an EMD-ANN-based method in which a noisy signal is first given to the EMD to be decomposed into a number of IMFs. Due to the different numbers of IMFs for each signal, these IMFs are then mapped to 13 and given as inputs to the neural network. The purpose of using a neural network in this study is that a neural network can act as a nonlinear filter and eliminate the need for thresholding. Moreover, the output of this network is the time-domain denoised signal.

The best neural network structure obtained in this study is a multilayer neural network with two hidden layers, the first and the second of which contain 25 and 20 neurons, respectively. Due to the mapping of IMFs into 13, the input of the neural network is also 13.

In this study, individual models were used for signal denoising at specific SNRs (0, 5, 10, 15, and 20dB) and different noises (white or pink). Since the noise in the hospital environment can be white or pink with any level of SNR [6], [21], it is necessary to use an algorithm to be able to denoise it without having to determine the type of noise and input SNR. The results obtained from the EMD-ANN and the study of denoised signal spectrograms showed that the combined model in both white and pink noises denoised the lung sound, whether bronchovesicular or vesicular, and improved the signal SNR similar to individual models while retaining much of the main content. Given that, there is no need to specify the input SNR in the combined model so that this model can be more valuable than the individual models.

The efficiency of the proposed method was further evaluated using four combined models with different noises (white and pink) with SNRs between -2 to 20dB with an increment by 1dB. The results showed that all four cases could improve the output SNR, which was greater in the white noise denoising. In addition, it can be concluded from the results that the combined model is capable of denoising the LS signal with the desired noise, which can make the use of this model much more desirable for use in real environments. This is because there are both white and pink noises in real environments, and the input SNRs can be different.

In the last step, the EMD-ANN was compared with the EMD-Custom method in terms of SNR improvement and the Fit value of the denoised signal. These results showed that although the EMD-Custom method was able to denoise the noisy signal at low SNRs, a large part of the main content of lung sound was removed as indicated by the output signal Fit value with SNR = 0 dB, and this method could only improve the output SNR. In addition, this method failed to improve the output SNR at high SNRs and has even reduced the quality of the output signal.

The results obtained in this paper showed that the combined model could be a reliable model for use in the real environment and eliminate the need to specify the predetermined parameters. This model will be able to denoise the noisy LS signal infected with white or pink noises with different SNRs between -2 to 20dB, regardless of any recorded area of the chest, while preserving the main content of the LS signal. Therefore, this model can be a good substitute for threshold-based noise reduction models.

There are also suggestions for future studies:

1- As mentioned above, one of the problems of EMD is the phenomenon of mode mixing, which can be solved using the EEMD algorithm.
2- The outstanding performance of this network in removing pink noise makes it possible to evaluate the performance of the proposed method to remove the heart sound from the lung sound signal.

## 8. Conflict of Interest

The authors have no conflict of interest relevant to this article.